\documentstyle[aps,multicol,amssymb,epsfig]{revtex}

\def\>{\rangle}
\def\<{\langle}

\newenvironment{figurehere}
   {\def\@captype{figure}}
   {}

\begin{document}
\title{A simple gate for linear optics quantum computing}
\author{Terry Rudolph and Jian-Wei Pan}
\address{Institut f\"ur Experimentalphysik, Universit\"at Wien, Boltzmanngasse 5, Vienna, Austria\\ } 

\maketitle

\begin{abstract}
We describe a simple scheme for implementing the non-linear sign gate of Knill, 
Laflamme and Milburn (Nature, {\bf 409}, 46-52, Jan. 4 (2001)) which forms the 
basis of an experiment underway at the University of Vienna. 
\end{abstract}

\begin{multicols}{2}
It was recently shown \cite{KLM} that efficient quantum computing is possible 
using only linear optics, single photon sources and single photon detectors. 
One of the fundamental gates of the proposed scheme is known as the non-linear 
sign (NLS) gate; it is a non-deterministic gate which implements the 
transformation (on states of photon occupation number) given by
\begin{equation}
\alpha|0\> +\beta |1\> + \gamma |2\> \rightarrow \alpha|0\> +\beta |1\> - \gamma 
|2\>,
\end{equation}
with probability of success 1/4. 

The scheme suggested in \cite{KLM} for implementing a NLS gate used a 
complicated interferometer requiring beamsplitters of variable reflectivity. We 
have found a scheme more amenable to an experimental demonstration of linear 
optics quantum computing, and it is depicted in Fig.~1. The input computational 
mode begins in a state of horizontal polarization, i.e. in general 
$|\psi_{in}\> = 
\alpha|0_H\> +\beta |1_H\> + \gamma |2_H\>$. This mode passes through a polarization rotator, of rotation angle 
$\sigma$. As with the proposal of \cite{KLM}, our scheme makes use of a single 
ancilla photon, in this case prepared with vertical polarization. The ancilla 
photon and computational mode are mixed at a polarizing beamsplitter (PBS). One 
output of the PBS goes to a detector $D_1$, and the gate's success is 
conditioned on no photons being detected at $D_1$. The other output mode of the 
PBS passes through a polarization rotator set to an angle $\theta$. This mode 
is subsequently subjected to a measurement, and the gate operation is 
successful if a single vertical photon is detected. This is indicated in the 
figure by the addition of a second PBS and detector $D_2$. In general the 
detector $D_2$ would need to be able to distinguish one from multiple photons, 
and such detectors are not readily available. However for the purposes of a 4 
photon coincidence experiment such multiple photon events are excluded by the 
conditioning process, and so a simple demonstration of an NLS gate can be 
performed using commonly available single photon detectors. 

If we take the transformation corresponding to a polarization rotator of angle 
$x$ to be $a_H^\dagger\rightarrow 
\cos x~a_H^\dagger+\sin x~a_V^\dagger$, $a_V^\dagger\rightarrow 
-\sin x~a_H^\dagger+\cos x~a_V^\dagger$, then the (unnormalized) state of the 
output mode, given the correct conditioning at $D_1,D_2$, is \cite{maple}
\begin{eqnarray}
|\psi_{out}\> &=& 
\alpha\cos\theta|0_H\>+\beta\cos\sigma\cos2\theta|1_H\> \nonumber\\
&&+\gamma\cos^2\sigma\cos\theta(1-3\sin^2\theta)|2_H\>. 
\end{eqnarray} 

\begin{figurehere}
\center{\epsfig{file=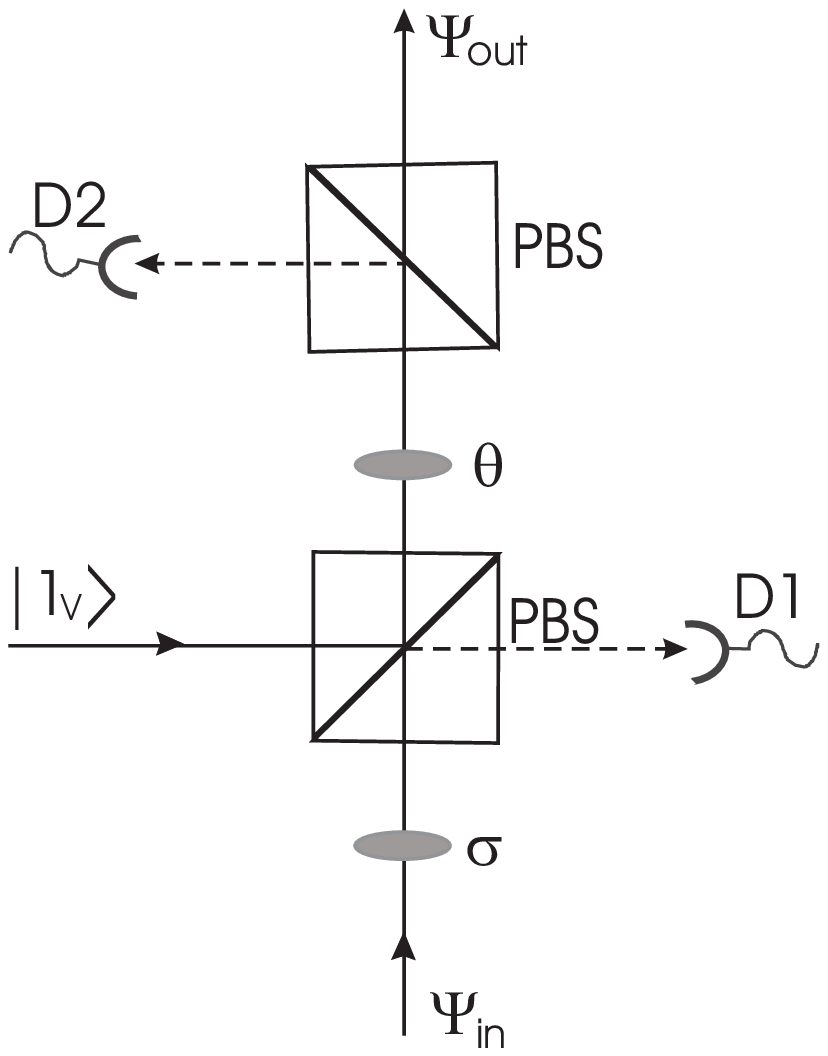,width=55mm,clip=}}
\\*
{\bf Fig.1.}{\it A simple scheme for implementing a NLS gate} 
\end{figurehere}
\vskip 0.3cm
Defining $A=\sqrt{21-7\sqrt{2}}/7$, one can easily show that choosing the 
angles $\theta=\arccos A\approx 61.5^o$ and 
$\sigma=\arccos((1-2\sqrt{2})A)\approx 150.5^o$, gives the desired 
transformation (1) with probability of success $A^2\approx 0.227$. This is 
slightly lower than the probability of success (0.25) for the scheme presented 
in \cite{KLM}, however the experimental simplification of our scheme is 
considerable. Furthermore the central result of \cite{KLM}, namely that this 
probability of success can be boosted arbitrarily close to 1 using only linear 
optics, single photon sources and single photon detectors, is unaffected by 
this small decrease in success probability of the NLS gate. 

This work was supported by the Austrian Science Foundation FWF, and the TMR 
programs of the European Union Project No. ERBFMRXCT960087. 
 
{\it Note added.} Recently a similar simplification of the NLS gate was 
presented \cite{tim} which shares the same probability of success as the one 
presented here. The scheme presented here enjoys the slight practical advantage 
of not requiring a beamsplitter with unequal reflectivity/transmittivity. 

\end{multicols}


\vskip -0.6cm
\begin{references}
\bibitem{KLM} E. Knill, R. Laflamme and G. Milburn, Nature, {\bf 409}, 46-52, Jan 4 
(2001).
\bibitem{maple} A Maple worksheet capable of calculating the state transformations associated with arbitrary 
linear optical interferometers (incorporating both spatial and polarization 
degrees of freedom) is available from T.R. on request. 
\bibitem{tim} T. Ralph {\it et. al.}, quant-ph/0108049.
\end{references}
\end{document}